\documentclass [11pt]{revtex4}
\usepackage {amsmath}

\begin{document}
\title{New exact solution of Dirac-Coulomb equation with exact boundary
condition}
\author{Ruida Chen}

\affiliation{Shenzhen Institute of Mathematics and Physics,
Shenzhen, 518028, China}


\begin{abstract}
It usually writes the boundary condition of the wave equation in the
Coulomb field as a rough form without considering the size of the
atomic nucleus. The rough expression brings on that the solutions of
the Klein-Gordon equation and the Dirac equation with the Coulomb
potential are divergent at the origin of the coordinates, also the
virtual energies, when the nuclear charges number  $Z > 137$,
meaning the original solutions do not satisfy the conditions for
determining solution. Any divergences of the wave functions also
imply that the probability density of the meson or the electron
would rapidly increase when they are closing to the atomic nucleus.
What it predicts is not a truth that the atom in ground state would
rapidly collapse to the neutron-like. We consider that the atomic
nucleus has definite radius and write the exact boundary condition
for the hydrogen and hydrogen-like atom, then newly solve the radial
Dirac-Coulomb equation and obtain a new exact solution without any
mathematical and physical difficulties. Unexpectedly, the $K$ value
constructed by Dirac is naturally written in the barrier width or
the equivalent radius of the atomic nucleus in solving the Dirac
equation with the exact boundary condition, and it is independent of
the quantum energy. Without any divergent wave function and the
virtual energies, we obtain a new formula of the energy levels that
is different from the Dirac formula of the energy levels in the
Coulomb field.
\end{abstract}


\maketitle

\section{Introduction}

The Dirac equation for the hydrogen atom has been treated in modern
mathematical physics textbooks\cite{Thaller:1992}. It is well known
that the Dirac equation succeed in many respects. The Dirac equation
is compatible with the theory of relativity, and it describes the
spin of the electron and its magnetic moment in a completely natural
way and so on. Especially, the distinguished Dirac formula of energy
levels in Coulomb field can explain the fine-structure of the
hydrogen atom. This is one of the important indicates of the
achievements of the Driac theory. However, the Dirac operator is not
apple-pie. According to the Ref.1, the main difficulty with a
quantum mechanical on-particle interpretation is the occurrence of
states with negative (kinetic) energy. Interaction may cause
transitions to negative energy states, so that there is no hope for
a stability of matter within that framework. On the other hand, we
have to be aware of the fact that a quantum mechanical
interpretation leads to inconsistencies if pushed too far. The
localization problem and the Klein paradox are still no clear
solution, even in quantum electrodynamics. Investigating the Dirac
equation, one should be not too far from Dirac's point of view:
``\ldots a book on the new physics, if not purely descriptive of
experimental work, must be essentially mathematical''. Here we start
with the above reference and take new attention to one of the
mathematical difficulties concealed in the Dirac equation with a
Coulomb potential, which have not been obtained any logical
mathematical explain, and we show finally that these paradoxes do
not exists actually, such as the virtual energies and the divergence
of all wave function on the ground state as $r \to 0$ in pure
Coulomb field. We show that all of this kind of difficulties is due
to the original incorrect mathematical methods for solving the
radial Dirac equation.

It is well known that, according to the boundary condition to solve
Schr\"{o}dinger equation\cite{Schrodinger:1926} for the hydrogen
atom one can naturally obtain the Bohr formula\cite{Bohr:1913} of
energy levels, this is the prominent sign that the quantum mechanics
is different from the classical mechanics. Considering the
relativistic effect, Dirac introduced his relativistic wave equation
for the single electron\cite{Dirac:1928} in 1928.
Darwin\cite{Darwin:1928} and Gordon\cite{Gordon:1928} first obtain
the exact solution of the Dirac equation with a Coulomb potential.
Biedenharm\cite{Biedenharm:1962}, Wong and
Yeh\cite{Wong:1982}\cite{Wong:1983}, Su\cite{Phys:1985} etc. also
constructed the different second Dirac equation and obtain the
different form of solutions. Nenciu\cite{Nenciu:1976}, Kalus and
W\"{u}st\cite{Klaus:1978} investigated the different construction
methods of self-adjoint extension of the Dirac operators with
coulomb potential, and it is also showed that the distinguished
self-adjoint extensions given by the two methods are identical. In
history, it was ineffectual to use the Klein-Gordon
equation\cite{Gordon:1926}\cite{Klein:1926}, to describe the
hydrogen atom because its eigenvalues of the quantum energy is
incompletely agrees accurately with the experimentally observed
hydrogen spectra. We investigated the mathematical foundation of
constructing all wave equations for the quantum system by the
numbers and found some mathematical problems that have been ignored
in the Dirac equation. We know that all wave equations for the
quantum system have not been strictly demonstrated in quantum
mechanics, but the Schr\"{o}dinger equation has not any mathematical
difficulty, it is regarded as a fundamental assumption and is
generally accepted. However the original solution of the Dirac
equation in Coulomb field is divergent at the origin of coordinate
and the energies for all atoms with nuclear charges number $Z > 137$
are the virtual numbers, all of these problems are actually serious
mistakes from the incorrect mathematical methods. If a solution of
some wave equation contains some mathematical or physical
difficulties even mistakes, it should imply some new laws that have
not been found\cite{Seevinck:2006}. The mathematical difficulties
concealed in the Dirac equation with the Coulomb potential should
imply some new conclusions that have not been found also.

For briefness to discuss the essential of the problems, we only
focus on the original exact solution of the Dirac equation for the
hydrogen atom in present paper. It is should be accentuated that the
boundary condition exerts decisive action in solving the wave
equation\cite{Ciftci:2005}\cite{Nakatsuji:2005}\cite{Gavrilov:1996}\cite{Vakarchuk:2005}\cite{Villalba:2005}\cite{Lujan:2006}\cite{Berkdemir:2006}\cite{Morsink:1991}.
However such important condition is not always attended in the
history of the relativistic quantum
mechanics\cite{Villalba:1994}\cite{Adame:1995}\cite{Horbatsch:1995}\cite{Phys:2002}\cite{Dong:2004}\cite{Marques:2005}\cite{Ritchie:2006}.
That the whole Dirac wave functions of the hydrogen and
hydrogen-like atom are divergent at the originof the coordinate
implies that the atom on the ground state would collapse to the
neutron-like. By all appearances, the deduction falls short of the
true. In fact, the original mathematical methods for solving the
radial Dirac equation are incorrect, although one can obtain the
fine-structure formula of the energy-levels. This is why we have
those mathematical difficulties. For finding the correct
eigensolution of the some differential equations that cannot be
optimized such as the radial Dirac equation with Coulomb, there is
some theorems which need ones to
prove\cite{Chen:2000}\cite{Chen:2003}. Any wave function must accord
with the conditions for determining solution and the physical
signification. The virtual energies and the divergence of the
original solution for the Dirac equation are due to the rough
boundary condition. Writing the exact boundary condition and solving
the Dirac equation outside the nucleus again, one can obtain a new
exact wave function which has no any mathematical paradox, including
the divergence of the wave function and the virtual energies.
Howvere, the new strict mathematical deduction indicates that the K
value constructed by Dirac is naturally written in the radius of the
hydrogen atom, one can find that the new formula of the energy
levels is no longer the Dirac formula, it is the correct result from
correct mathematical method for solving the redial Dirac equation.

\section{Rough boundary condition and divergence of Driac
function}

Using the Schr\"{o}dinger equation to describe the quantum system of
the hydrogen atom, it usually does not consider the size of the
atomic nucleus. The boundary condition of the atom with nuclear
charges number Z is written as the rough form
\begin{equation}
\label{eq1} R\left( {r \to 0} \right) \ne \pm \infty ,\quad R\left(
{r \to \infty } \right) = 0,\quad - \infty < R\left( {0 < r < \infty
} \right) < \infty
\end{equation}

\noindent where $R$ is the radial wave function. Solving the
Schr\"{o}dinger equation by using this rough boundary condition, the
Bohr formula of the energy levels is naturally obtained, it is one
of the most consummate parts in quantum mechanics. In Dirac theory,
because of absence for considering size of the atomic nucleus, one
still uses the above rough boundary condition to solve the radial
Dirac equation and obtains the distinguished Dirac formula of the
energy levels in the Coulomb field\cite{Dirac:1958}
\begin{equation}
\label{eq2} E = \frac{mc^2}{\sqrt {1 + {\alpha ^2} \mathord{\left/
{\vphantom {{\alpha ^2} {\left( {n_r + \sqrt {K^2 - Z^2\alpha ^2} }
\right)^2}}} \right. \kern-\nulldelimiterspace} {\left( {n_r + \sqrt
{K^2 - Z^2\alpha ^2} } \right)^2}} }
\end{equation}

\noindent where $K = \pm 1,\;\pm 2,\;\pm 3,\; \cdots $, constructed
by Driac, $n_r = 0,1,2, \cdots $, and $\alpha $ is the fine
structure constant. It sticks out a mile, for the ground state or
says S-state also, $n = 0,\;K = \pm 1$, when Z>137, the energies of
the system must be virtual numbers. This is pure mathematical
problems. We don't think that those explain which departure
mathematics too far are correct.

On the other hand, the corresponding Dirac wave function takes the
form with two components
\begin{equation}
\label{eq3} R\left( r \right) = \left( {{\begin{array}{*{20}c}
 {e^{ - ar}\sum\limits_{\nu = 0}^n {b_\nu } \left( {ar} \right)^{\sqrt {K^2 - Z^2\alpha ^2} + \nu - 1}} \hfill \\
 {e^{ - ar}\sum\limits_{\nu = 0}^n {d_\nu } \left( {ar} \right)^{\sqrt {K^2 - Z^2\alpha ^2} + \nu - 1}} \hfill \\
\end{array} }} \right)
\end{equation}

\noindent where $a = {\sqrt {m^2c^4 - E^2} } \mathord{\left/
{\vphantom {{\sqrt {m^2c^4 - E^2} } {\hbar c}}} \right.
\kern-\nulldelimiterspace} {\hbar c}$ and the coefficients of the
polynomial satisfy the system of the recursive relations
\begin{equation}
\label{eq4}
\begin{array}{l}
 \frac{1}{\lambda }b_{\nu - 1} + d_{\nu - 1} - Z\alpha b_\nu - \left( {K + \sqrt {K^2 - Z^2\alpha ^2} + \nu } \right)d_\nu = 0 \\
 b_{\nu - 1} + \lambda d_{\nu - 1} + \left( {K - \sqrt {K^2 - Z^2\alpha ^2} - \nu } \right)b_\nu + Z\alpha d_\nu = 0 \\
 \end{array}
\end{equation}

\noindent where $\lambda = \sqrt {\frac{mc^2 - E}{mc^2 + E}} $.
However the wave function for the S state is divergent, as $K = 1$,
whatever $n_r $ takes any value, the expression (\ref{eq3}) becomes
\begin{equation}
\label{eq5} R\left( {r \to 0} \right) = \mathop {\lim }\limits_{r
\to 0} \left( {{\begin{array}{*{20}c}
 {\left( {ar} \right)^{\sqrt {K^2 - Z^2\alpha ^2} - 1} + \sum\limits_{\nu = 1}^{n_r } {b_\nu } \left( {ar} \right)^{\sqrt {K^2 - Z^2\alpha ^2} + \nu - 1}} \hfill \\
 {\left( {ar} \right)^{\sqrt {K^2 - Z^2\alpha ^2} - 1} + \sum\limits_{\nu = 1}^{n_r } {d_\nu } \left( {ar} \right)^{\sqrt {K^2 - Z^2\alpha ^2} + \nu - 1}} \hfill \\
\end{array} }} \right)e^{ - ar} = \left( {{\begin{array}{*{20}c}
 \infty \hfill \\
 \infty \hfill \\
\end{array} }} \right)
\end{equation}

\noindent this is a typical mathematical difficulty, which the
solution balance out the condition for determining solution of the
wave equation.

Any mathematical difficulty in physical theory must lead to some
deductions that contravene the order of nature. When looking from a
physical point of view, the divergence of the Dirac wave function
for S state implies that the probability density of the electron
around the nucleus rapidly increases as it close to the atomic
nucleus. The probability density from the wave function with two
components is defined as
\begin{equation}
\label{eq6} \rho \left( {r,t} \right) = R^ + \left( {r,t}
\right)R\left( {r,t} \right)
\end{equation}

According to (\ref{eq3}), we have
\begin{equation}
\label{eq7} R^ + = e^{ - ar}\left( {{\begin{array}{*{20}c}
 {\sum\limits_{\nu = 0}^n {b_\nu \left( {ar} \right)^{\nu + \sqrt {K^2 - Z^2\alpha ^2} - 1}} \quad } \hfill & {\sum\limits_{\nu = 0}^n {d_\nu \left( {ar} \right)^{\nu + \sqrt {K^2 - Z^2\alpha ^2} - 1}} } \hfill \\
\end{array} }} \right)
\end{equation}

\noindent In this case the radial probability density of the
electron for the relativistic hydrogen is as follows
\begin{equation}
\label{eq8} \rho  = \left[ {e^{ - ar} \sum\limits_{\nu  = 0}^n
{b_\nu  \left( {ar} \right)^{\nu  + \sqrt {K^2  - Z^2 \alpha ^2 }  -
1} } } \right]^2  + \left[ {e^{ - ar} \sum\limits_{\nu  = 0}^n
{d_\nu  \left( {ar} \right)^{\nu  + \sqrt {K^2  - Z^2 \alpha ^2 }  -
1} } } \right]^2
\end{equation}

\noindent For S-state which implies$K = 1$, as $r \to 0$, the above
formula becomes
\begin{equation}
\label{eq9} \mathop {\lim \rho }\limits_{r \to 0} = \infty
\end{equation}

\noindent what this result predicts should be that the hydrogen and
hydrogen-like atom in the ground state must rapidly collapse to the
neutron-like. However the fact is not thusness. That is to say, the
original solution of the Dirac equation for the hydrogen and
hydrogen-like atom neither agrees with the mathematical principle
nor agrees with the physical signification. Unexpectedly, such
divergence was defined as so-called ``mild
divergence''\cite{Taub:1949}\cite{Yennie:1954}\cite{Bethe:1957}so
that hardly might one open out its actual meaning, and the correct
deduction have been buried. We know that the Klein-Gordon for the
meson without spin has the same divergence, but the Klein-Gordon
divergence in the Coulomb field can be eliminated by the suitable
mathematical method. Only one demonstrate some new theorems for
finding the eigensolutions set of some differential equations with
the variable coefficient can find the correct eigenvalues set of the
corresponding wave equations (Ref.31 and Ref.32).

Using a cut-off procedure for the potential that is similar to the
case of considering an extended nucleus to blench the divergence
should be independent of the exact solution for the Dirac equation
with the Coulomb potential, and it oppresses the exact solution of
the wave equation. For the exact solution, why coming forth the
mathematical difficulty such as the expression (\ref{eq5}) and
(\ref{eq9}) and the virtual energies is that the size of the nucleus
of the hydrogen atom is not considered in the rough boundary
condition (\ref{eq1}), and the nuclear are regarded as the point in
geometrical meaning. The point in geometrical meaning falls short of
the actual case of the atomic nucleus. In fact, the necessary of the
normalizable wave function of the hydrogen atom has been discussed
home and widely in some modern physics textbook\cite{Greiner:2000}.
Now one should consider the actual size of the atomic nucleus to
rewrite the boundary condition then find the eigensolution of the
Dirac equation for the hydrogen and hydrogen-like atom.

\section{Exact boundary condition and new solution of Dirac
equation}

Boundary conditions for any wave equations are written out basing on
the structure of the physical model and distributing character of
the physical quantity. Consider two basic facts, one is that the
atomic nucleus has definite size, we suppose its equivalent radius
or barrier width is $\delta $. Another is that the electron does not
enter the inside of the atomic nucleus, and does not collide to and
rub with the atomic nucleus. In this way, any wave equation that
describes the atom has the same exact boundary condition
\begin{equation}
\label{eq10} R\left( {r \le \delta } \right) \ne  \pm \infty ,\quad
R\left( {r \to \infty } \right) = 0,\quad  - \infty  < R\left(
{\delta  < r < \infty } \right) < \infty
\end{equation}

\noindent one would recover the Bohr formula of the energy levels if
uses this exact condition to solve the Schr\"{o}dinger equation for
the hydrogen atom. It is well known that the Schr\"{o}dinger
equation is very consummate in mathematics. However the new formula
of the energy levels coming from the Dirac equation with the exact
boundary condition is not as exact as the Dirac formula with the
rough boundary condition and the divergence for the wave function.
Of course, one would obtain the satisfying formula that is as exact
as the distinguished Dirac formula when considering the spin-orbit
coupling in the Dirac equation. It should be not one and only choice
that for explaining the fine structure of the hydrogen atom one
still uses the rough boundary condition for the Dirac equation and
still blench those mathematical and physical difficulty such as the
divergence and the virtual energies. One note that the boundary
conditions are related to the self-adjointness of the operator, also
one of methods to obviate the divergence of the Dirac function was
given by Deck, Amar and Fralick\cite{Deck:2005}.

Now use the exact boundary condition (\ref{eq10}) to solve the
radial Dirac equation of the hydrogen atom. It usually introduces a
mathematical transformation
\begin{equation}
\label{eq11} R = \left( {{\begin{array}{*{20}c}
 {{F\left( r \right)} \mathord{\left/ {\vphantom {{F\left( r \right)} r}} \right. \kern-\nulldelimiterspace} r} \hfill \\
 {{G\left( r \right)} \mathord{\left/ {\vphantom {{G\left( r \right)} r}} \right. \kern-\nulldelimiterspace} r} \hfill \\
\end{array} }} \right)
\end{equation}

\noindent and translate the radial Dirac equation for the hydrogen
atom\cite{Schiff:1968}\cite{Bjorken:1965}$^{ }$
\begin{equation}
\label{eq12} \left[ {c\hat {p}_r \left( {{\begin{array}{*{20}c}
 {\;0} \hfill & i \hfill \\
 { - i} \hfill & 0 \hfill \\
\end{array} }} \right) - \frac{\hbar cK}{r}\left( {{\begin{array}{*{20}c}
 0 \hfill & {\;1} \hfill \\
 1 \hfill & {\;0} \hfill \\
\end{array} }} \right) + mc^2\left( {{\begin{array}{*{20}c}
 1 \hfill & {\;0} \hfill \\
 0 \hfill & { - 1} \hfill \\
\end{array} }} \right)} \right]R = \left( {E + \frac{e^2}{4\pi \varepsilon _0 r}} \right)R
\end{equation}

\noindent into the following form$^{ }$
\begin{equation}
\label{eq13}
\begin{array}{l}
 \left( {\frac{{E - mc^2 }}{{\hbar c}} + \frac{{Z\alpha }}{r}} \right)F + \left( {\frac{k}{r} + \frac{d}{{dr}}} \right)G = 0 \\
 \left( {\frac{{E + mc^2 }}{{\hbar c}} + \frac{{Z\alpha }}{r}} \right)G + \left( {\frac{k}{r} - \frac{d}{{dr}}} \right)F = 0 \\
 \end{array}
\end{equation}

\noindent where $K = \pm 1,\pm 2, \cdots $ Considering the exact
boundary condition (\ref{eq10}), introduce the transform
\begin{equation}
\label{eq14} \xi  = r - \delta \quad \left( {\xi  \ge 0} \right)
\end{equation}

\noindent the boundary (\ref{eq10}) can be overwritten as follows
\begin{equation}
\label{eq15} R\left( {\xi \to 0} \right) \ne \pm \infty ,\quad
R\left( {\xi \to \infty } \right) = 0,\quad - \infty < R\left( {0 <
\xi < \infty } \right) < \infty
\end{equation}

\noindent then $r = \xi + \delta $, substituting it into
(\ref{eq13}), one obtains
\begin{equation}
\label{eq16}
\begin{array}{l}
 \left( {\frac{E - mc^2}{\hbar c} + \frac{\alpha }{\xi + \delta }} \right)F + \left( {\frac{K}{\xi + \delta } + \frac{d}{d\xi }} \right)G = 0 \\
 \left( {\frac{E + mc^2}{\hbar c} + \frac{\alpha }{\xi + \delta }} \right)G + \left( {\frac{K}{\xi + \delta } - \frac{d}{d\xi }} \right)F = 0 \\
 \end{array}
\end{equation}

\noindent Because $F \to 0$, $G \to 0$ as $\xi \to \infty $, the
system of differential equations (\ref{eq16}) has the formal
solutions with weight function of asymptotic solutions
\begin{equation}
\label{eq17} F = e^{ - a\xi }f\left( \xi \right),\mbox{ }G = e^{ -
a\xi }g\left( \xi \right)
\end{equation}

\noindent Substituting (\ref{eq17}) into equation (\ref{eq16}), one
then obtains
\begin{equation}
\label{eq18}
\begin{array}{l}
 \left[ {\frac{E - mc^2}{\hbar c}\left( {\xi + \delta } \right) + \alpha } \right]f + \left( {\xi + \delta } \right)\frac{dg}{d\xi } + \left[ {K - a\left( {\xi + \delta } \right)} \right]g = 0 \\
 \left[ {\frac{E + mc^2}{\hbar c}\left( {\xi + \delta } \right) + \alpha } \right]g - \left( {\xi + \delta } \right)\frac{df}{d\xi } + \left[ {K + \left( {a\xi + \delta } \right)} \right]f = 0 \\
 \end{array}
\end{equation}

\noindent the Eigensolutions of equations (\ref{eq18}) correspond to
quantum energy are two interrupted series£¬which the number of terms
is determined by the eigenvalues.

In order to find the general series solutions for equations
(\ref{eq18}), it is assumed that the formal solutions are
\begin{equation}
\label{eq19} f\left( \xi \right) = \sum\limits_{\nu = 0}^\infty
{b_\nu \xi ^{\sigma + \nu }} ,\quad g\left( \xi \right) =
\sum\limits_{\nu = 0}^\infty {d_\nu \xi ^{\sigma + \nu }}
\end{equation}

\noindent Substituting into equations (\ref{eq18}), one obtains the
linear system of recursive relations
\begin{equation}
\label{eq20}
\begin{array}{l}
 \sum\limits_{\nu  = 0}^\infty  {\left[ \begin{array}{l}
 \frac{{E - mc^2 }}{{\hbar c}}b_{\nu  - 1}  + \frac{{E - mc^2 }}{{\hbar c}}\delta b_\nu   + \alpha b_\nu   + Kd_\nu   \\
  + \left( {\sigma  + \nu } \right)d_\nu   + \delta \left( {\sigma  + \nu  + 1} \right)d_{\nu  + 1}  - ad_{\nu  - 1}  - \delta ad_\nu   \\
 \end{array} \right]\xi ^{\sigma  + \nu } }  = 0 \\
 \sum\limits_{\nu  = 0}^\infty  {\left[ \begin{array}{l}
 \frac{{E + mc^2 }}{{\hbar c}}d_{\nu  - 1}  + \frac{{E + mc^2 }}{{\hbar c}}\delta d_\nu   + \alpha d_\nu   + Kb_\nu   \\
  - \left( {\sigma  + \nu } \right)b_\nu   - \delta \left( {\sigma  + \nu  + 1} \right)b_{\nu  + 1}  + ab_{\nu  - 1}  + \delta ab_\nu   \\
 \end{array} \right]\xi ^{\sigma  + \nu } }  = 0 \\
 \end{array}
\end{equation}

\noindent hence the coefficient of the power series satisfy the
following system of recursive relations
\begin{equation}
\label{eq21}
\begin{array}{l}
 \frac{{E - mc^2 }}{{\hbar c}}b_{\nu  - 1}  + \left( {\frac{{E - mc^2 }}{{\hbar c}}\delta  + \alpha } \right)b_\nu   - ad_{\nu  - 1}  \\
  + \delta \left( {\sigma  + \nu  + 1} \right)d_{\nu  + 1}  + \left( {K + \sigma  + \nu  - \delta a} \right)d_\nu   = 0 \\
 \frac{{E + mc^2 }}{{\hbar c}}d_{\nu  - 1}  + \left( {\frac{{E + mc^2 }}{{\hbar c}}\delta  + \alpha } \right)d_\nu   + ab_{\nu  - 1}  \\
  - \delta \left( {\sigma  + \nu  + 1} \right)b_{\nu  + 1}  + \left( {K - \sigma  - \nu  + \delta a} \right)b_\nu   = 0 \\
 \end{array}
\end{equation}

\noindent Corresponding to $\nu = - 1$ the indicial equations are
given that $\delta \sigma b_0 = 0$ and $\delta \sigma d_0 = 0$.
Because $\delta \ne 0$, $b_0 \ne 0$ and $d_0 \ne 0$, one obtains
\begin{equation}
\label{eq22} \sigma = 0
\end{equation}

\noindent so that the wave functions satisfy the boundary condition
at $r \to \delta $ namely $\xi \to 0$, the above equations reduce to
\begin{equation}
\label{eq23}
\begin{array}{l}
 \frac{E - mc^2}{\hbar c}b_{\nu - 1} + \left( {\frac{E - mc^2}{\hbar c}\delta + \alpha } \right)b_\nu - ad_{\nu - 1} + \delta \left( {\nu + 1} \right)d_{\nu + 1} + \left( {K + \nu - \delta a} \right)d_\nu = 0 \\
 \frac{E + mc^2}{\hbar c}d_{\nu - 1} + \left( {\frac{E + mc^2}{\hbar c}\delta + \alpha } \right)d_\nu + ab_{\nu - 1} - \delta \left( {\nu + 1} \right)b_{\nu + 1} + \left( {K - \nu + \delta a} \right)b_\nu = 0 \\
 \end{array}
\end{equation}

\noindent Respectively evaluate for $\nu = 0,1,2, \cdots ,n_r $,
$b_{n_r + 1} = d_{n_r + 1} = 0$, make use of that $b_{ - 2} = d_{ -
2} = 0$ and $b_{ - 1} = d_{ - 1} = 0$, equations (\ref{eq23}) give
\begin{equation}
\label{eq24}
\begin{array}{l}
 \left( {\frac{E - mc^2}{\hbar c}\delta + \alpha } \right)b_0 + \left( {K - \delta a} \right)d_0 + \delta d_1 = 0 \\
 \left( {K + \delta a} \right)b_0 + \left( {\frac{E + mc^2}{\hbar c}\delta + \alpha } \right)d_0 - \delta b_1 = 0 \\
 \frac{E - mc^2}{\hbar c}b_0 + \left( {\frac{E - mc^2}{\hbar c}\delta + \alpha } \right)b_1 - ad_0 + \left( {K + 1 - \delta a} \right)d_1 + 2\delta d_2 = 0 \\
 ab_0 + \left( {K - 1 + \delta a} \right)b_1 + \frac{E + mc^2}{\hbar c}d_0 + \left( {\frac{E + mc^2}{\hbar c}\delta + \alpha } \right)d_1 - 2\delta b_2 = 0 \\
 \cdots \cdots \cdots \cdots \cdots \cdots \cdots \cdots \cdots \cdots \cdots \cdots \cdots \cdots \cdots \cdots \cdots \cdots \cdots \cdots \cdots \\
 \frac{E - mc^2}{\hbar c}b_{n_r - 1} + \left( {\frac{E - mc^2}{\hbar c}\delta + \alpha } \right)b_{n_r } - ad_{n_r - 1} + \left( {K + n_r - \delta a} \right)d_{n_r } = 0 \\
 ab_{n_r - 1} + \left( {K - n_r + \delta a} \right)b_{n_r } + \frac{E + mc^2}{\hbar c}d_{n_r - 1} + \left( {\frac{E + mc^2}{\hbar c}\delta + \alpha } \right)d_{n_r } = 0 \\
 \frac{E - mc^2}{\hbar c}b_{n_r } - ad_{n_r } = 0 \\
 ab_{n_r } + \frac{E + mc^2}{\hbar c}d_{n_r } = 0 \\
 \end{array}
\end{equation}

\noindent The last two formulas are linearly dependent. Use ${\left(
{E + mc^2} \right)} \mathord{\left/ {\vphantom {{\left( {E + mc^2}
\right)} {\hbar c}}} \right. \kern-\nulldelimiterspace} {\hbar c}$
to multiply the third formula from bottom and use ${\sqrt {m^2c^4 -
E^2} } \mathord{\left/ {\vphantom {{\sqrt {m^2c^4 - E^2} } {\hbar
c}}} \right. \kern-\nulldelimiterspace} {\hbar c}$ to multiply the
fourth formula from bottom, and then add the two new formulas, it is
given as follows
\begin{equation}
\label{eq25}
\begin{array}{l}
 \left[ {\alpha \left( {E + mc^2 } \right) + \left( {K - n_r } \right)\sqrt {m^2 c^4  - E^2 } } \right]b_{n_r }  \\
  + \left[ {\left( {K + n_r } \right)\left( {E + mc^2 } \right) + \alpha \sqrt {m^2 c^4  - E^2 } } \right]d_{n_r }  = 0 \\
 \end{array}
\end{equation}

\noindent Substituting for the second formula from the bottom, one
will obtain a new formula of the energy levels for the hydrogen atom
\begin{equation}
\label{eq26} E = \frac{mc^2}{\sqrt {1 + \left( {\frac{\alpha }{n_r
}} \right)^2} }\mbox{ }\left( {n_r = 1,2,3, \cdots } \right)
\end{equation}

\noindent One can strictly demonstrate that $n_r  \ge 1$. It is
different from the Dirac formula of the energy levels for the
hydrogen atom. This result is the inevitable deduction of the Dirac
equation with the exact boundary condition for the hydrogen atom.
With the exact boundary condition (\ref{eq10}) and the new formula
of the energy levels (\ref{eq26}), all of the corresponding wave
functions satisfy the boundary conditions and there is not any
virtual energies.

According to (\ref{eq11}), (\ref{eq14}), (\ref{eq16}), (\ref{eq19}),
(\ref{eq23}), the whole wave function with the exact boundary
condition is as follows
\begin{equation}
\label{eq27}
 R = \left( {\begin{array}{*{20}c}
   {\frac{{e^{ - a\xi } }}{{\xi  + \delta }}\sum\limits_{\nu  = 1}^{n_r } {b_\nu  } \xi ^\nu  }  \\
   {\frac{{e^{ - a\xi } }}{{\xi  + \delta }}\sum\limits_{\nu  = 1}^{n_r } {d_\nu  } \xi ^\nu  }  \\
\end{array}} \right),\quad \left( {n_r  \ge 1} \right)
\end{equation}

\noindent the coefficients of corresponding polynomial are
determined by the system of recursive relations (\ref{eq23}). All
appearance, at the boundary of the hydrogen atom
\begin{equation}
\label{eq28} \mathop {\lim }\limits_{\xi \to 0} R = \left(
{{\begin{array}{*{20}c}
 {\mbox{Constant}} \hfill \\
 {\mbox{Constant}} \hfill \\
\end{array} }} \right),\quad \mathop {\lim }\limits_{\xi \to \infty } R = \left( {{\begin{array}{*{20}c}
 0 \hfill \\
 0 \hfill \\
\end{array} }} \right)
\end{equation}

\noindent Make use of the definition (\ref{eq6}), the probability
density of the electron appearing outside the nucleus of the
hydrogen atom takes the form
\begin{equation}
\label{eq29} \rho = \left( {\frac{e^{ - a\xi }}{\xi + \delta
}\sum\limits_{\nu = 0}^{n_r } {b_\nu } \xi ^\nu } \right)^2 + \left(
{\frac{e^{ - a\xi }}{\xi + \delta }\sum\limits_{\nu = 0}^{n_r }
{d_\nu } \xi ^\nu } \right)^2
\end{equation}

\noindent homoplastically, one has
\begin{equation}
\label{eq30} \mathop {\lim }\limits_{\xi \to 0} \rho =
\mbox{Constant},\quad \mathop {\lim }\limits_{\xi \to \infty } \rho
= 0
\end{equation}

\section{Conclusions}

In this paper we expatiated that the divergences of the Dirac
function and the virtual energies of the Dirac formula of the energy
levels for the hydrogen and hydrogen-like atom are due to the
traditional rough boundary condition. By using the exact boundary
condition one can obtain a new solution of the Dirac equation in the
Coulomb field. The new solution without any mathematical difficulty
gives a new formula of the energy levels which is different from the
distinguished Dirac formula. One can find that in the new solution
the $K$ values constructed by Dirac have been written in the radius
of the atomic nucleus and they are~independent of the new formula of
the energy levels. Only when looking from point of view for the
formula of the energy levels, the new formula is not as exact as the
Driac formula. However, considering the spin-orbit coupling or some
new potential parameters, one will obtain the exacter formula.

It is well known that the Dirac equation succeed in many important
prognostics$^{ }$\cite{Buchanan:2005}. Dirac's equation has an
infinite number of solutions with negative energies. It leads to
discover the existence of the positron and implies an unexpected
relativistic interaction between an electron's translational motion
and spin, which leads to a violent oscillation of the particle at
very high frequencies and over distances of roughly one Compton
wavelength. For Dirac's electrons, zitterbewegung takes place
whenever the electron wave function includes both positive and
negative energy components. It takes both sets of states to build up
an arbitrary electronic state. The distillate of quantum mechanics
is to naturally obtain the formula of the energy levels for the
bound state by solving the wave equation with the boundary
condition. Disclosing the mathematical and physical difficulty
concealing in the Dirac equation with the Coulomb potential does not
imply to negative the Dirac theory. Our works wishes to vindicate
the mathematical logic for the wave equation very precise, at least
to eliminate the divergence by using the exact boundary condition to
replace the rough boundary condition.

No matter which wave equation is used to describe the quantum system
of the bound statge, their eigenvalues set and eigensolutions set
must be in agreement with the uniqueness, and the solution must
accord with the conditions for the exact solution but not any
approximate solution such as cut off potential. In principle, we
cannot immolate the mathematical rule to obtain a formula only for
agreeing with the experimentally observed hydrogen spectra. Actually
the divergence of the original solution for the Dirac equation in
the Coulomb field is nonexistent. The Dirac equation is more and
more widely applied for various models. One should note that using
different mathematical methods and different boundary condition to
solve the differential equation will obtain different
results\cite{Dong:2003}\cite{Alhaidari:2004}. The rough boundary
condition for the Dirac equation with the Coulomb potential brings
on so mathematical contradictions. We need to revise all incorrect
mathematical methods and search the correct mathematical methods to
find the correct and exact solution of the various wave equations.
One can find that the classical solution of the plane transverse
electromagnetic mode of the Maxwell equation is also
incorrect\cite{Chen:2001}, because of the incorrect mathematical
method for solving the wave equation. For quantum mechanics, only
the exact solution of the wave equation can we know if the
differential wave equation has the eigenvalues connecting the energy
levels. Further research should explore the exacter formula of the
energy levels for the hydrogen and hydrogen-like atom by using the
exact boundary condition that accords with the structure of the
atoms. Here it should be pointed out that the Schr\"{o}dinger
equation and the Klein-Gordon equation in the Coulomb field with the
exact boundary condition have the same corresponding formula of the
energy levels.

\end{document}